\documentclass[%
reprint,
superscriptaddress,
amsmath,amssymb,
aps,
]{revtex4-1}

\usepackage{dcolumn}
\usepackage{bm}

\usepackage[utf8x]{inputenc}
\usepackage[normalem]{ulem}
\usepackage[british]{babel}
\usepackage{amsmath,amsfonts,amssymb,graphicx,color,amsbsy}
\usepackage{gensymb}

\usepackage{geometry}            
\usepackage{natbib}
\usepackage{mathtools}
\usepackage[version=3]{mhchem}
\usepackage{latexsym} 
\numberwithin{equation}{section}
\usepackage{import}
\newcommand{\lapla}{\nabla ^2}
\newcommand{\eps}{\epsilon}

\newcommand{\nn}{\nonumber}

\begin{document}

\preprint{APS/123-QED}

\title{Localized Instabilities and Spinodal Decomposition in Driven Systems in the Presence of Elasticity}

\author{Esteban Meca}
\affiliation{%
Weierstrass Institute, Mohrenstr. 39, 10117 Berlin, Germany 
}%
\author{Andreas M\"unch}
\affiliation{%
Mathematical Institute, University of Oxford, Andrew Wiles Building, Woodstock Road, Oxford, OX2 6GG UK
}%
\author{Barbara Wagner}
\email{Corresponding author: wagnerb@wias-berlin.de}
\affiliation{%
Weierstrass Institute, Mohrenstr. 39, 10117 Berlin, Germany 
}%

\date{\today}

\begin{abstract}
We study numerically and analytically the instabilities associated with phase separation in a solid layer on which an external material flux is imposed. The first instability is localized within a boundary layer at the exposed free surface by a process akin to spinodal decomposition. In the limiting static case, when there is no material flux, the coherent spinodal decomposition is recovered. In the present problem stability analysis of the time-dependent and non-uniform base states as well as numerical simulations of the full governing equations are used to establish the dependence of the wavelength and onset of the instability on parameter settings and its transient nature as the patterns eventually coarsen into a flat moving front. The second instability is related to the Mullins-Sekerka instability in the presence of elasticity and arises at the moving front between the two phases when the flux is reversed. Stability analyses of the full model and the corresponding sharp-interface model are carried out and compared.
Our results demonstrate how interface and bulk instabilities can be analysed within the same framework which allows to identify and distinguish each of them clearly.
The relevance for a detailed understanding of both instabilities and their interconnections in a realistic setting are demonstrated for a system of equations modelling the lithiation/delithiation processes within the context of Lithium ion batteries. 
\begin{description}
 \item[PACS numbers]
68.43.Jk, 81.10.Aj, 81.15.Aa 
 \end{description}
\end{abstract}

\maketitle



\section{Introduction}\label{sec:intro}

Localized instabilities in phase transformations in non-equilibrium systems have been investigated for a long time. Possibly the most well-known example is the Mullins-Sekerka interfacial instability of solidifying systems  \cite{Mullins1963,Mullins1964}, which has also been studied in the presence of elasticity for coherent interfaces \cite{Leo1989,Onuki1991}. Similar interaction of a diffusional instability with elasticity have also been intensely studied and are well-known as the Asaro-Tiller-Grinfeld  instability \cite{Asaro1972,Grinfeld1986,Srolovitz1989} resulting from the competition of surface diffusion and stress relaxation. 

Spinodal decomposition in the bulk is another common phenomenon that can be understood as an instability in phase-separating systems. The celebrated theory of Cahn and Hilliard \cite{Cahn1958,Cahn1961} gave a foundation for the understanding of this phenomenon as a bulk instability. Through spinodal decomposition a system phase-separates, e.g. in a binary system  regions with a higher concentration of solute are instantaneously created. While it is well known that the Cahn-Hilliard approach has limitations \cite{Gunton1983}, the approach remains very useful for early stages of spinodal decomposition, allowing the incorporation of additional effects, that may facilitate or suppress the instability. Most common in material science are effects of elasticity, anisotropy \cite{Cahn1961,Cahn1962}, or surface induced spinodal decomposition \cite{Puri2005,Hennessy2014}. In addition, the coupling of spinodal decomposition with elasticity in thin films has also received much attention in connection with defects \cite{Ipatova1993,Leonard1997,Leonard1998a} and also with surface instabilities, in particular the Asaro-Tiller-Grinfeld  instability \cite{Leonard1998}. 

Recently, the study of the interaction of spinodal decomposition and elasticity has intensified due to newly discovered localization effects. Phase-field simulations of thin films have shown that the instability tends to be localized first near the free surface of the film \cite{Seol2003, Wise2005, Zhen2006}. A similar result had been reported by Ipatova et al. \cite{Ipatova1993}, who showed that due to elastic effects spinodal decomposition can be localized exponentially close to the surface in elastically anisotropic epitaxial films. 
Moreover, this exponentially-localized surface mode can become unstable even when the bulk is stable \cite{Tang2012}. The concentrations at which this mode is unstable lay between the classical (chemical) spinodal and the spinodal modified by elastic effects (coherent spinodal). This type of localized instabilities seem to underly a number of fundamental processes such as the stability of grain boundaries in phase-separating systems that is currently receiving much attention  \cite{Geslin2015,Xu2016}, where an understanding of localized instabilities in the presence of coherency strain is of capital importance. 

In addition, novel technological applications of spinodal decomposition in thin films are emerging, ranging from a means to engineer the mechanical properties of a thin film \cite{Rogstroem2012} to a technique of obtaining optically active silicon nanoparticles \cite{Roussel2013}. The initial motivation of the present study concerns an instability during the  lithiation/delithiation process of phase-changing electrodes used for example in Lithium-ion batteries \cite{Meca2016a}.  It has long been known that electrode materials such as \ce{LiFePO4} undergo phase separation when lithiated or delithiated, and this has been studied using extensions of the Cahn-Hilliard model \cite{Cogswell2012,Bazant2013}. Some promising high capacity electrode materials such as amorphous silicon (a-\ce{Si}), is known to also undergo two-phase lithiation \cite{Wang2013}. However doubts remain regarding the mechanical properties, which have been tested for instance in the experiments of Sethuraman et al. \cite{Sethuraman2010}. Recently, it has been conjectured that phase separation should be taken into account to explain the observed mechanical properties \cite{Meca2016a}, and a simplified model for the experimental setup in \cite{Sethuraman2010} was developed. The model describes a thin layer of a-\ce{Si} that has been grown on a crystalline substrate and is lithiated from the free surface. The increasing concentration of lithium in the layer causes the volume of the layer to increase, and when the concentration is high enough the system undergoes phase separation and a highly lithiated phase is created near the free surface, showing a periodic structure for some values of the system parameters. As  the pattern moves into the amorphous layer under continued flux it coarsens into a flat front that moves into the layer. If the flux is reversed this front undergoes an interfacial  instability. 
Since these instabilities emerge within non-uniform driven systems it is necessary to investigate the connection of  localization of instabilities near the free surface with interfacial instabilities using a unified framework. 

In order to study this instability we use a viscous Cahn-Hilliard model \cite{Novick1988} to model phase separation, and couple the dynamics of the concentration with elasticity using what is usually referred to as the Larch\'e-Cahn prescription \cite{Larche1982,Onuki1989,Fratzl1999}. We also use the sharp-interface limit of this model \cite{Meca2016b}, which is valid once phase separation has taken place. Comparing the results of the phase-field model with the sharp-interface model allows us to on the one hand validate the stability calculation and on the other hand show how the localization of the instability occurs in the phase-field model.

We solve numerically the model in two dimensions and study the development of an instability related with spinodal decomposition, but in the presence of a driving flux that further confines it to the free surface. We study the instability by computing the eigenvalues and eigenvectors of the linearized sytem for a laterally unbounded layer, in the "frozen-time" or adiabatic approximation \cite{Meca2010,Hennessy2013}. Additionally we study the stability of a receding front using the same technique and relate it with the stability of the front as described by the sharp-interface model.

In Section \ref{TheModel} we give a summary description of the model used, and in Section \ref{Stability} we study the linearized model. In Section \ref{Numerics} we give a brief description of the numerics, and in Sections \ref{Results} and \ref{Conclusion} we present the numerical results of the direct simulation and the different stability calculations and discuss them.

\section{The Model\label{TheModel}}

In this section we introduce the model used. This is a model for the lithiation of a layer of amorphous silicon that has been described elsewhere\cite{Meca2016a}, and hence it is not our goal to describe in detail the derivation of the model.

\begin{figure}[htbp]
\includegraphics[width=\columnwidth]{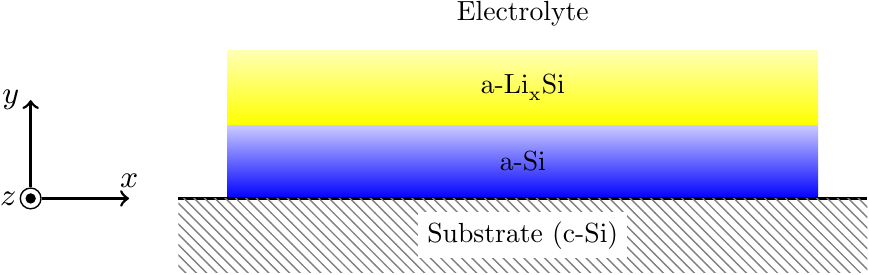}
\caption{Scheme of the amorphous silicon layer.}
\label{fig:scheme}
\end{figure}

In our description, we have $c$, a dimensionless concentration of solute (the local molar fraction of lithium) inside of a layer of amorphous silicon (see Fig.~\ref{fig:scheme}). We assume that the deformations are small, and hence we can use linear elasticity. The strain tensor $\epsilon_{ij}$ is defined as
\begin{equation}
\epsilon_{ij}=\frac{1}{2}\left(\partial_j u_i + \partial_i u_j\right),
\end{equation}
in terms of the deformation $\mathbf{u}$, with the indices $1\le i,j\le 3$. We will use nevertheless the plane strain approximation, and hence $u_z=0$ and all derivatives with respect to $z$ cancel. The elastic energy is defined as
 \begin{equation}
W=\frac{1}{2}C_{ijkl}\left(\epsilon_{ij}-\epsilon^0_{ij}\right)\left(\epsilon_{kl}-\epsilon^0_{kl}\right) \label{eq:elasticenergy},
\end{equation}
where the summation is implied, and $C_{ijkl}$ is the fourth order elasticity tensor. Since the material of interest is amorphous we will assume it to be fully isotropic. The stress-free strain or eigenstrain is defined as $\epsilon^0_{ij}=\alpha h(c)\delta_{ij}$, where the constant $\alpha$ is the maximum stress-free strain and $h(c)$ is an interpolating monotone function such that $h(0)=0$ and $h(1)=1$. The stress is defined as follows:
\begin{align}
\sigma_{ij}&=\frac{\partial W}{\partial \epsilon_{ij}}=C_{ijkl}\left(\epsilon_{kl}-\epsilon^0_{kl}\right)\nonumber\\ &=\frac{E(c)}{1+\nu}\left[\epsilon_{ij}-\epsilon^0_{ij}+\frac{\nu}{1-2\nu}(\epsilon_{kk}-\epsilon^0_{kk})\delta_{ij}\right],
\end{align}
where $E(c)$ is Young's modulus (which depends on the concentration) and $\nu$ is Poisson's ratio. For the problem at hand, we assume that Young's modulus depends on the concentration, with the extreme values being for pure amorphous silicon $E(0)=E_{\ce{Si}}$ and for fully lithiated a-\ce{Si} $E(1)=E_{\ce{Li _x Si}}$. The value of $\nu$ is not expected to show a strong dependence with respect to the concentration, in accordance with Shenoy et al. \cite{Shenoy2010}.

The total free energy of the layer reads:
\begin{equation}
\mathcal{F}=\int_{\Omega}\left(\frac{1}{2}\gamma\varepsilon \left|\nabla c\right|^2+\frac{\gamma}{\varepsilon} f(c)+W(\epsilon_{ij},c)\right)dxdy,
\end{equation}
where the homogeneous free energy density $f(c)=c^2(1-c)^2/4$
and $W(\epsilon_{ij},c)$ is the elastic energy density as defined in Eq.~\eqref{eq:elasticenergy}.  The constant $\gamma$ carries the dimensions of energy over length and the parameter $\epsilon$ is proportional to the interface thickness. The chemical potential reads
\begin{equation}
\mu=\frac{\delta\mathcal{F}}{\delta c}=-\gamma\varepsilon\nabla^2 c+\frac{\gamma}{\varepsilon}f'(c)+\partial_c W(\epsilon_{ij},c), \label{eq:chempot}
\end{equation}
and we use the following equation for the dynamics of the concentration:
\begin{equation}
\partial_t c =M \nabla^2\left(\mu+\chi\varepsilon\partial_t c\right), \label{eq:CH}
\end{equation}
where M is a constant mobility and $\chi$ is the viscosity parameter. Eq.~\ref{eq:CH} would have the familiar form of the Cahn-Hilliard equation but for the last term, the viscous term  \cite{Novick1988}. While this term is not commonly used in Cahn-Hilliard-like models, it is important as it captures part of the non-equilibrium kinetics of the interface. Gurtin \cite{Gurtin1996} showed that such a term appears naturally when introducing $\partial_t c$ in the list of constitutive variables, and it has been shown to guarantee a positive entropy production at the interface in the sharp-interface limit\cite{Dreyer2015}. The chosen scaling from that term with $\varepsilon$ follows similarly from the sharp-interface limit of this model (see \cite{Meca2016b}). Eqs.~\ref{eq:CH}, \ref{eq:chempot} together with the mechanical equilibrium condition
\begin{equation}
\partial_j\sigma_{ij}=0,
\end{equation}
are the equations that define the dynamics of our system.

In order to nondimensionalize the system, we introduce a lengthscale $H_0$ that corresponds to the height of the layer in the absence of lithium. The resulting system has the following form (see \cite{Meca2016a} for the details of the scalings):

\begin{subequations}
\label{eq:ndimeqs}
\begin{align}
\partial _t c&=\nabla ^2 \left(\mu+\varepsilon \beta\, \partial _t c\right),\label{eq:mu}\\
\mu&=-\varepsilon\nabla^2 c+ \frac{1}{\varepsilon}f'(c)+\xi\,\partial_c W\!\left(\epsilon_{ij},c\right),\\
\partial_j\sigma_{ij}&=0, \label{eq:stress_ndim}\\
\sigma_{ij}&=2{G} \left(\epsilon_{ij}-\epsilon^{0}_{ij}\right)+\frac{2\nu}{1-2\nu}{G}\left(\epsilon_{kk}-\epsilon^{0}_{kk}\right)\delta_{ij}, \label{eq:nsc}
\end{align}
where the constitutive laws for the
nondimensional shear modulus $G=E(c)/E_{\ce{Si}}$ 
and stress-free strain $\epsilon^{0}_{ij}$ are specified as
\begin{align*}
{G}&=1+g(c)\left(\frac{E_{\ce{Li_xSi}}}{E_{\ce{Si}}}-1\right),\quad
\epsilon^{0}_{ij}=h(c)\delta_{ij}, 
\end{align*}
and the derivative of the nondimensional elastic energy takes the 
form
\begin{align}
&\partial_c W(\epsilon_{ij},c)=\nonumber\\
&\frac{(1-\nu)G'}{1-2\nu}\left(\partial_1u_1^2+\partial_2u_2^2\right)
+\frac{1}{2}G'\left(\partial_1u_2+\partial_2u_1\right)^2\nonumber\\
&+\frac{2\nu G'}{1-2\nu}\partial_1u_1\partial_2u_2
-\frac{2(1+\nu)}{1-2\nu}(h(c)G)'\nabla\cdot \mathbf{u}\nonumber\\
&+\frac{3 (1+\nu)}{1-2\nu}\left(h(c)^2G\right)'. 
\end{align}
Here, $h(c)$ and $g(c)$ are interpolating functions such that $g(0)=h(0)=0$ and $g(1)=h(1)=1$.  
For the boundaries in contact with the substrate, we will take a no-flux/no-deformation boundary condition:
\begin{equation}\label{eq:bcs_subs_ndim}
\mathbf{u}=0, \qquad
\mathbf{n}\cdot\nabla c=0, \qquad  
\mathbf{n}\cdot\nabla \mu=0, 
\end{equation}
where $\mathbf{n}$ is the normal vector to the surface.
In the case of the boundaries in contact with the electrolyte, we take a no-traction boundary condition and, following \cite{Burch2009}, assume a consistent no-flux condition for $c$ (also known as {\it variational boundary condition}), together with a constant flux boundary condition
\begin{equation}\label{eq:bcs_elec_ndim}
\mathbf{\sigma}\cdot\mathbf{n}=0,\qquad
\mathbf{n}\cdot\nabla c=0, \qquad
\mathbf{n}\cdot\nabla \mu=K\!\left(\mu\right)=F.
\end{equation}
\end{subequations}
The function $K(\mu)$, which in our case is simply equal to the constant $F$, is in general a nonlinear function of the chemical potential, and relates the absorption into the layer with the outer electrical potential. While the phenomenological Butler-Volmer relation is commonly used (see e.g. \cite{Zeng2014}) but there exist more rigorous aproaches \cite{Bazant2013}. In our case, the constant $F$ corresponds to a galvanostatic lithiation regime.

The problem depends on the following non-dimensional groups:
\begin{equation}
\beta=\frac{\chi M}{H_0},\quad
F=\frac{F_r H_0^2}{M \gamma},\quad
\xi=\frac{H_0 E_{\ce{Si}}\alpha^2}{2(1+\nu)\gamma},
\end{equation}
where $F_r$ is the dimensional flux. The previous parameters, together with the elastic ratio $E_{\ce{Li_xSi}}/E_{\ce{Si}}$, Poisson's ratio $\nu$ and $\epsilon$ are the complete set of non-dimensional parameters. Note that $\xi$ is the ratio of elastic to interfacial energies.

For the numerical simulations we have used $E_{\ce{Li_xSi}}/E_{\ce{Si}}=0.44$ and $\nu=0.25$, in accordance with the calculations form Shenoy et al. \cite{Shenoy2010}.

\section{Stability Analysis \label{Stability}}

In this section, we consider the case of a laterally unbounded layer that is delimited by $y=0$ and $y=1$. We derive the system of equations that a linear perturbation about a basis solution given by a one-dimensional displacement and concentration profile fulfils. Specifically, we assume a basis solution of the form
\begin{align}
&u_x(x,y,t)=0,\qquad u_y(x,y,t)=u_{y,0}(y,t),\nonumber\\ 
&c(x,y,t)=c_{y,0}(y,t).
\end{align}

If we perturb this solution slightly we obtain:
\begin{subequations}
\label{eq:ansatz}
\begin{align}
u_x(x,y,t)&=\delta\, u_{x,1}(y,t)e^{ikx},\\
u_y(x,y,t)&=u_{y,0}(y,t)+\delta\, u_{y,1}(y,t)e^{ikx},\\
c(x,y,t)&=c_{0}(y,t)+\delta\, c_{1}(y,t)e^{ikx},
\end{align}
\end{subequations}
where $\delta$ is a formal expansion parameter.

We introduce this ansatz into Eqs.~\eqref{eq:ndimeqs} and obtain for the $O(\delta)$ terms of the stress:
\begin{subequations}
\begin{align}
\sigma_{xx,1}&=\frac{2G(c_0)}{1-2\nu}\left[ik(1-\nu)u_{x,1}+\nu u'_{y,1}\right.\nonumber\\
&\left.-(1+\nu)h'(c_0)c_1\right]-2G'(c_0)\frac{1+\nu}{1-\nu}h(c_0)c_1,\\
\sigma_{yy,1}&=\frac{2G(c_0)}{1-2\nu}\left[(1-\nu)u'_{y,1}+ik\nu u_{x,1}\right .\nonumber\\
&\left . -(1+\nu)h'(c_0)c_1\right],\\
\sigma_{xy,1}&=G(c_0)\left(iku_{y,1}+u'_{x,1}\right),
\end{align}
\end{subequations}
where the prime symbol denotes derivative either with respect to the argument (as in $G$ or $h$) or derivative with respect to $y$, in $c_1$, $u_{x,1}$ and $u_{y,1}$.

The stress balance equations \eqref{eq:stress_ndim} read
\begin{subequations}
\begin{align}
ik\sigma_{xx,1}+G'(c_0)\partial_{y}c_0\left(iku_{y,1}+u'_{x,1}\right) &
\nonumber\\
+G(c_0)\left(iku'_{y,1}+u''_{x,1}\right)=0 &,\\
ik\sigma_{xy,1}+\frac{2G'(c_0)}{1-2\nu}\partial_{y}c_0\left[(1-\nu)u'_{y,1} \right.& \nonumber\\
\left.+ik\nu u_{x,1}-(1+\nu)h'(c_0)c_1\right] & \nonumber\\
+\frac{2G(c_0)}{1-2\nu}\left[(1-\nu)u''_{y,1}+ik\nu u'_{x,1}\right. & \nonumber\\
\left. -(1+\nu)\partial_{y}c_0h''(c_0)c_1-(1+\nu)h'(c_0)c'_1\right]=0 &.
\end{align}
\label{eq:perturbstress}
\end{subequations}

And the concentration balance equation \eqref{eq:mu} has the following form
\begin{subequations}
\begin{align}
\partial_t c_1 &= \mathrm{D}\left(\mu_1+\varepsilon \beta\, \partial _t c_1\right)\\
\mu_1&=-\varepsilon \mathrm{D}c_1+\frac{1}{\varepsilon}f''(c_0)c_1 \nonumber\\  &\qquad +2\xi\frac{1+\nu}{1-\nu}\left[(Gh^2)''c_1+\frac{1+\nu}{1-2\nu}Gh'^2c_1\right.\nonumber\\
&\qquad\left.-Gh'\frac{1-\nu}{1-2\nu}(iku_{x,1}+u'_{y,1})-i k G'hu_{x,1}\right]
\end{align}
\label{eq:perturbconc}
\end{subequations}
where $\mathrm{D}\vcentcolon=\partial_y^2-k^2$.

The boundary conditions at $y=0$ are
\begin{equation}
\partial_yc_1=0, \qquad\partial_y\mu_1=0, \qquad\mathbf{u}_1=0.\label{eq:BCsyeq0}
\end{equation}

and at $y=1$
\begin{equation}
\partial_yc_1=0, \qquad\partial_y\mu_1=0, \qquad\boldsymbol{\sigma}_1\cdot\mathbf{n}=0.\label{eq:BCsyeq1}
\end{equation}

The last condition on stress can be replaced by the following conditions in terms of the displacements

\begin{subequations}
\begin{align}
u'_{y,1}&=-ik\frac{\nu}{1-\nu} u_{x,1}+\frac{1+\nu}{1-\nu}h'(c_0)c_1,\\
u'_{x,1}&=-iku_{y,1}.
\end{align}
\label{eq:BCsyeq1_ex}
\end{subequations}

Eqs.~\eqref{eq:perturbstress}, \eqref{eq:perturbconc} with boundary condtions \eqref{eq:BCsyeq0}, \eqref{eq:BCsyeq1} and \eqref{eq:BCsyeq1_ex} can be turned into a real system of equations with the change $i u_x\rightarrow \tilde{u}_x$. We adopt this convention in the following.

In order to study the stability we adopt the "frozen time" approximation \cite{Hennessy2013}, also called sometimes adiabatic approximation \cite{Meca2010}. In this approximation, the time dependence of the coefficients of the equation is not considered, and only the time dependence of the perturbation is taken into account to solve the equation. In our case,  this means that the time dependence that enters in Eqs.~\eqref{eq:perturbstress} and \eqref{eq:perturbconc} through $c_0$ is ignored.

The solution of Eqs..~\eqref{eq:perturbstress} and \eqref{eq:perturbconc} can then be written as a generalized eigenvalue problem, where the eigenvalues correspond to the growth rate of the perturbation. This generalized eigenvalue problem can then be solved numerically.

\section{Numerics\label{Numerics}}

The equations \eqref{eq:ndimeqs} have been solved in one and two dimensions using an non-linear adaptive multigrid algorithm for the spatial part \cite{Wise2007} and a Crank-Nicolson time stepping scheme. This algorithm is implemented in the solver BSAM.

In order to solve the linearized system for the perturbations given by Eqs..~\eqref{eq:perturbstress} and \eqref{eq:perturbconc} the equations are discretized using a pseudospectral method, Chebyshev collocation. The resulting system can be casted as a generalized eigenvalue problem, with the eigenvalues being the growth rate of the perturbations. The system is then solved by Arnoldi's method using ARPACK routines as implemented in Matlab. The output of the BSAM solver is fed into the pseudospectral method by means of a resampling and interpolation, and the resolution is increased to ensure convergence and avoid the problems inherent to the resampling. We use four levels of refinement on the adaptive multigrid, which corresponds to $\Delta x=1.95\times 10^{-3}$ at its smallest, and use a number of Chebyshev collocation points that is enough to resolve this.

Regarding the specific choice of the auxiliary interpolating functions, we choose $g(c)=c$, implying a linear decrease of Young's modulus with $c$ and $h(c)=c$, which follows from Vegard's law for the eigenstrain. The non-dimensional parameters $\beta$ and $\xi$ are varied across several orders of magnitude to observe their effect, since they are not know experimentally for the model system proposed. The value of the flux parameter $k$ is in principle adjustable experimentally, and we have picked it to be $k=4.0$. Finally the interface width is selected to be $\varepsilon=0.005$ except where indicated. See also \cite{Meca2016a} for a comprehensive exploration of the effect of the different parameters in this system.

\section{Results\label{Results}}

\subsection{Two-dimensional simulations}

We have studied the behaviour of a layer with a rectangular cross-section. This study can be performed in two dimensions through the plane-strain approximation. The layer has a ratio of height to width of $1/4$, is clamped on the substrate below it and has a no-flux on all sides except the upper one, on which the flux is applied (see Fig.~\ref{fig:scheme}). 

\begin{figure*}[htbp]
\def\svgwidth{\textwidth}
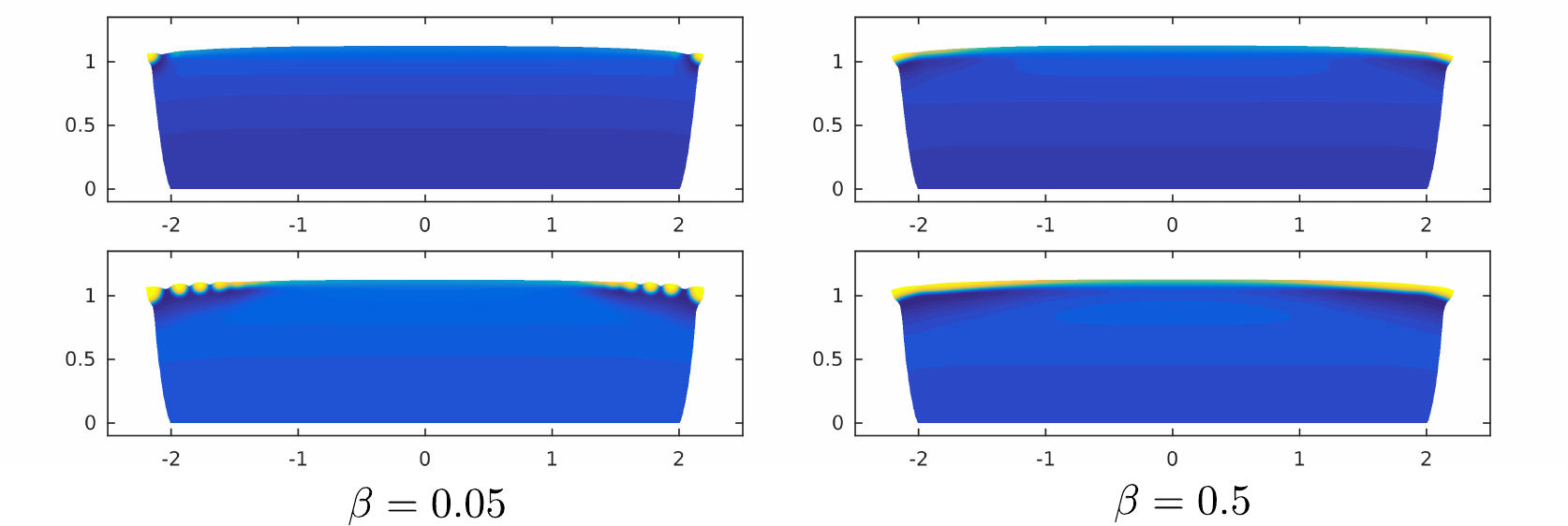
\caption{Onset of the instability for $\xi=0.1$. For $\beta=0.05$ the periodic structure near the corners is clearly visible, as it is its evolution from the corner spot (times, from top to bottom $t=0.0378$, $t=0.0380$). For $\beta=0.5$ this periodicity is no longer present, and instead phase separation occurs smoothly, starting likewise from the corners (times, from top to bottom $t=0.0388$,$t=0.0390$)}
\label{fig:compbeta}
\end{figure*}

The initial condition corresponds to a completely depleted undeformed layer, on which a constant flux is applied. For the system at hand, this corresponds to a galvanostatic lithiation of the electrode. The initially rectangular domain deforms then on the top side, since it is where most of the lithium is accumulated. This accumulation eventually leads to phase separation on the upper part of the layer, see Fig.~\ref{fig:compbeta}.

Phase separation occurs in different ways depending on the values of the parameters. For a small value of the kinetic parameter $\beta=0.05$, the instability begins with a small pearl of the lithiated phase formed near the corners of the layer, which spreads then towards the center of the upper side following a periodic pattern. The instability begins in a corner due to our particular geometric choice, since it is there where the stress is the smallest and hence phase separation in incentivated by its smaller energy cost.

For the higher value of $\beta=0.5$, we see in Fig.~\ref{fig:compbeta} that this periodic behaviour is notoriously absent, and the onset of the instability is slightly delayed. This delay due to kinetic effects is to be expected on general grounds (see \cite{Hillert2003} and also \cite{Meca2016a} for the application to this system), and the reason for the instability to lose its periodicity is discussed below in connection with the stability analysis.

\begin{figure*}[htbp]
\def\svgwidth{\textwidth}
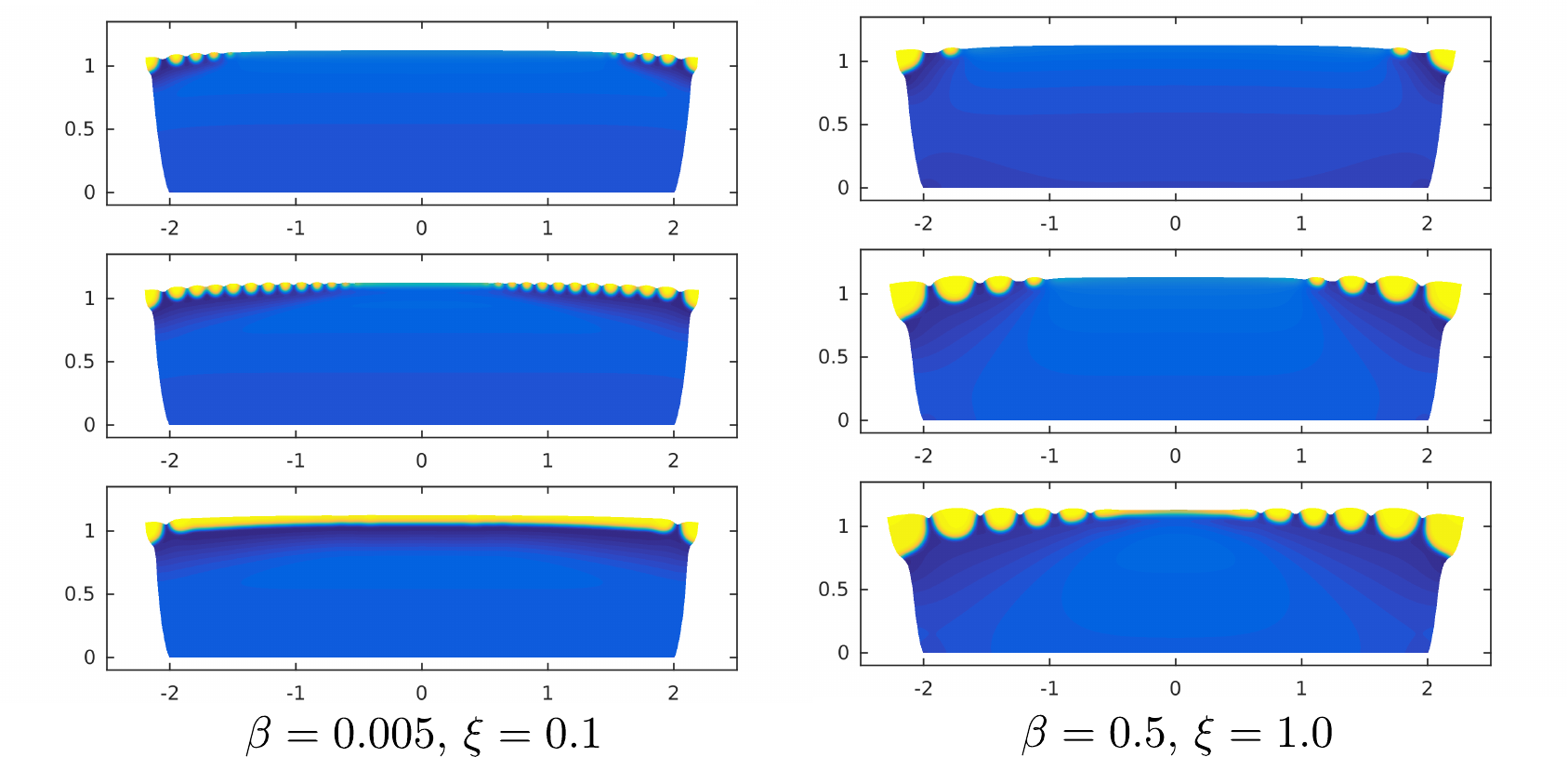
\caption{Effect of a higher value of $\xi$ at the onset of the instability. The instability develops with a mostly well-defined periodicity for $\xi=0.1$ and $\beta=0.005$, but it is very short lived as the initial lithiated "pearls" coarsen almost immediately (times, from top to bottom $t=0.0378$, $t=0.0380$, $t=0.0383$). For $\xi=1.0$ and $\beta=0.5$ the instability develops in a much slower fashion and gives rise to  lithiated pearls of a greater size that persist in time (times, top to bottom $t=0.0393$, $t=0.0413$, $t=0.0423$).}
\label{fig:compxi}
\end{figure*}

Increasing the value of $\xi$ similarly delays the onset of the instability. Again, this is to be expected since increasing $\xi$ lowers the position of the coherent spinodal. Higher values of $\xi$ bring nevertheless a curious interplay of effects (see Fig.~\ref{fig:compxi})

For a small value of the kinetic parameter $\beta=0.005$ and $\xi=0.1$ the instability develops but coarsens almost instantly. For larger values of $\xi$ this is not the case. Even in the $\beta=0.5$ case that did not show any signs of instability we observe for $\xi=1.0$ a periodic instability with a smaller spatial frequency.  A large value of $\xi$ delays phase transition and hence, when it occurs, a large volume of lithiated silicon is generated near the corners. At the interface larger values of the stress are present, and hence the associated elastic energy discourages the phase transition near the interface, and hence the wavelength of the instability must be larger. At the same time, the size of the initial grain is much larger for the $\xi=1.0$ case than for the $\xi=0.1$ case, thus we anticipate the importance of the nonlinear effects to explain this effect.

\subsection{Linear stability analysis of the Localized modes}

In this section we study the stability of the laterally unbounded system. The solution of the one-dimensional problem is introduced into the system formed by Eqs.~\eqref{eq:perturbstress} and \eqref{eq:perturbconc}, and we solve the associated eigenvalue problem as a function of time.

\begin{figure}[htbp] 
   \centering
   \includegraphics[width=0.95\columnwidth]{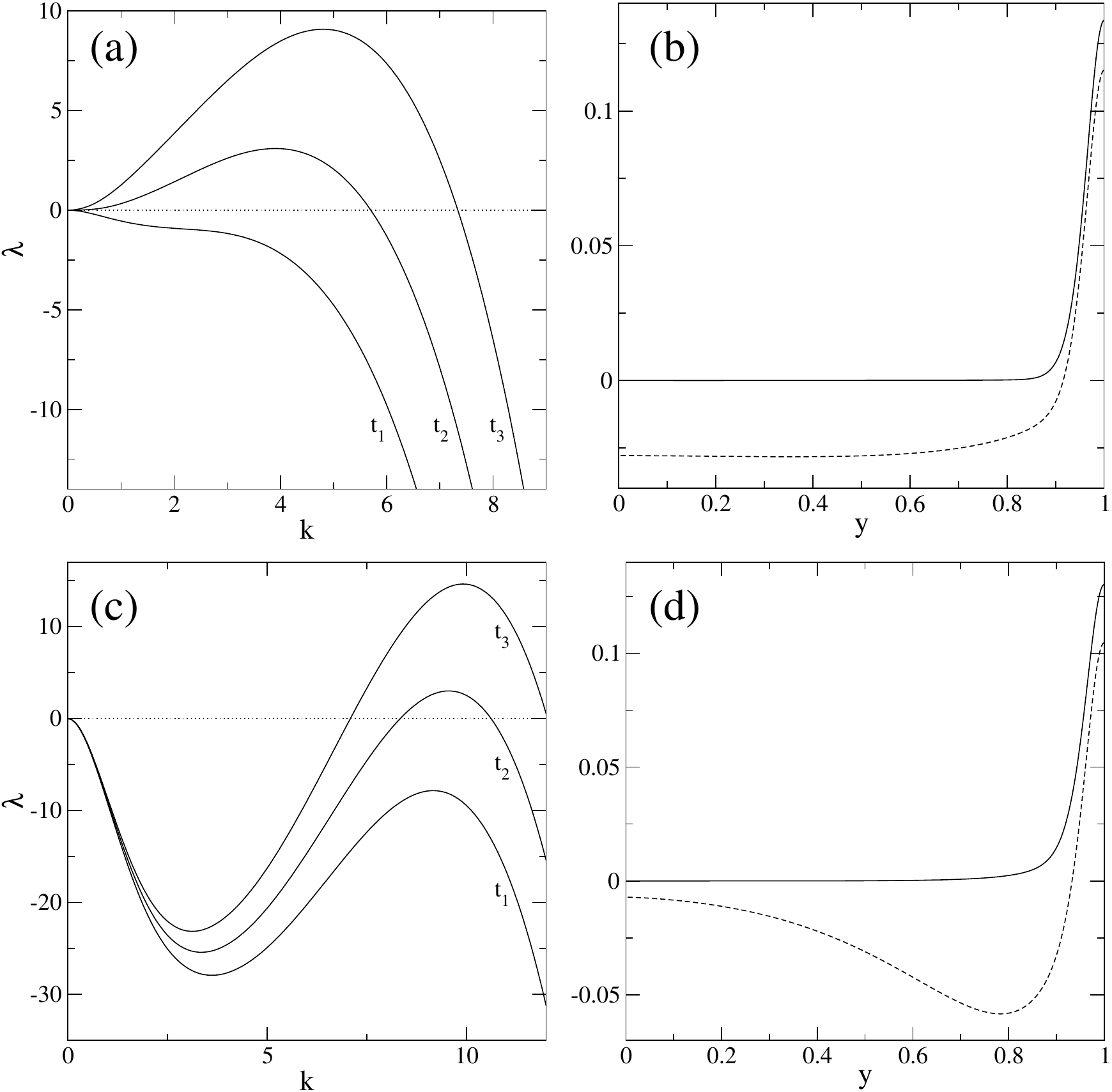} 
   \caption{Instability for $\beta=0.005$ and $\xi=0.1$ (a-b) and $\xi=1.0$ (c-d). (a) Growth rate as a function of wavenumber for the times $t_1=0.0367$, $t_2=0.0367125$, and $t_3=0.036725$. (b) Most unstable eigenvectors at the onset,  $t=0.0367125$ and $\mathrm{k}=3.9$. The solid and the dashed lines correspond respectively to the most unstable and the second most unstable eigenvectors. (c) Growth rate as a function of wavenumber for the times $t_1=0.0391$, $t_2=0.0391125$, and $t_3=0.039125$. (d) Most unstable eigenvectors at the onset,  $t=0.0391125$ and $\mathrm{k}=9.55$. The solid and the dashed lines correspond again to the most unstable and the second most unstable eigenvectors.}
   \label{fig:disprel}
\end{figure}

The dispersion relation is obtained by computing the largest eigenvalue as a function of the wavenumber $\mathrm{k}$. The results show that the dispersion relation is zero at $\mathrm{k}=0$ in the vicinity of the onset and, as opposed to spinodal decomposition, the instability starts at a finite value of $\mathrm{k}$. While this behaviour is not evident for $\xi=0.1$ (see Fig.~\ref{fig:disprel}a), it can clearly be observed for $\xi=1.0$ (Fig.~\ref{fig:disprel}c), thus showing that this is an effect that clearly stems from the coupling with elasticity. The value of $\mathrm{k}$ at which the growth rate is at a maximum ($\mathrm{k}_{max}$) increases steadily as the system becomes more unstable (see Fig.~\ref{fig:disprel}), in a behaviour similar to that found for the dispersion relation associated with spinodal decomposition (see e.g. Ref.~\cite{Gunton1983}). 

In addition to the dispersion relation we have also computed the most unstable eigenvectors for $\xi=0.1$ and $\xi=1.0$ at the onset (see Fig.~\ref{fig:disprel}). Results show a ver strong confinement near the surface, with a width of the layer mostly independent of $\xi$. We see nevertheless that the second most-unstable eigenvector, which is not localized, is different for $\xi=0.1$ (Fig.~\ref{fig:disprel}b) and $\xi=1.0$ (Fig.~\ref{fig:disprel}d), where it strongly undershoots.

The previous localized instability can be compared with that from Tang et al.  \cite{Tang2012}. For a constant concentration basis state, we obtain a good agreement with their results for a large enough size of the system, despite the differences in the treatment of elasticity. Nevertheless, note that the similarity between the leading eigenvector in Figs.~\ref{fig:disprel}b and \ref{fig:disprel}d shows that the confinement of the eigenvectors  is an effect mostly related with the imposed flux, whereas the confinement in Ref.~\cite{Tang2012} is a consequence of elasticity. For a small enough value of the flux $F$ we would recover an almost-flat concentration profile and then the scenario discussed in Ref.~\cite{Tang2012} would be the relevant one.

\begin{figure}[htbp] 
   \centering   \includegraphics[width=\columnwidth]{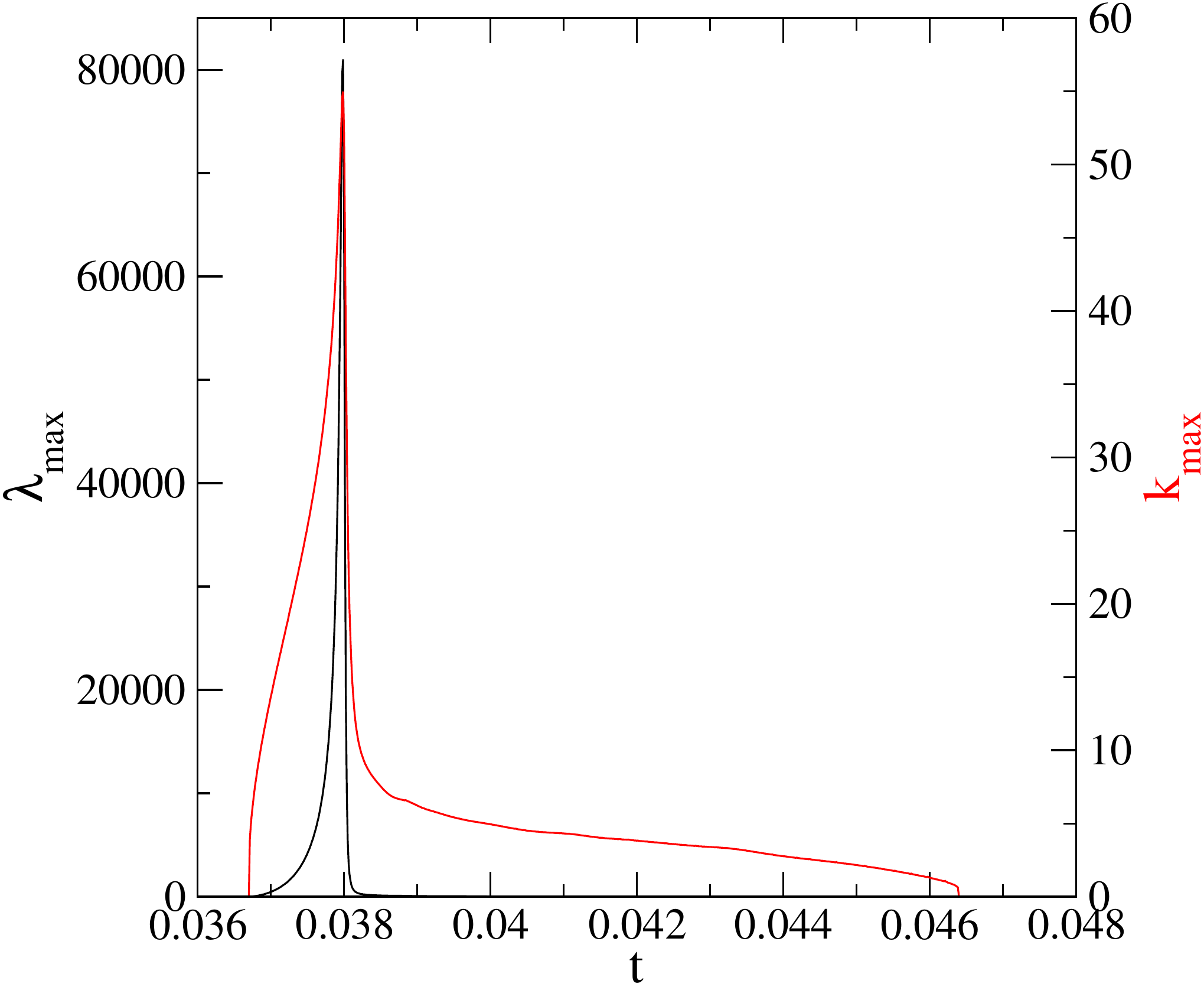} 
   \caption{Development of the instability for $\beta=0.005$ and $\xi=0.1$. Evolution of $\lambda_{max}$ and $\mathrm{k}_{max}$ with time.}
   \label{fig:evollambdamax}
\end{figure}

In Fig.~\ref{fig:evollambdamax} the evolution of the instability is visualized by computing the maximum value of the growth rate ($\lambda _{max}$) as a function of time. The instability develops very quickly, reaching large values of $\mathrm{k}_{max}$ and $\lambda _{max}$, only to decay even at a faster pace. After decaying, the instability settles for a short time into a long-wave mode with a very small growth rate, which is unlikely to be observed.

The comparison of the results on Figs.~\ref{fig:disprel} and \ref{fig:evollambdamax} for $\beta=0.005$, $\xi=0.1$ with those shown on Fig.~\ref{fig:compxi} show that the peak of the instability corresponds indeed to the instability found in the two-dimensional simulations. The instability peaks  at $t\approx 0.038$ with a value of $\mathrm{k}_{max}\approx 55$, which results in a wavelength of about $0.11$ units of length, which close to the one observed near the central areas in Fig.~\ref{fig:compxi}.

\begin{figure*}[htbp] 
   \centering
   \includegraphics[width=\textwidth]{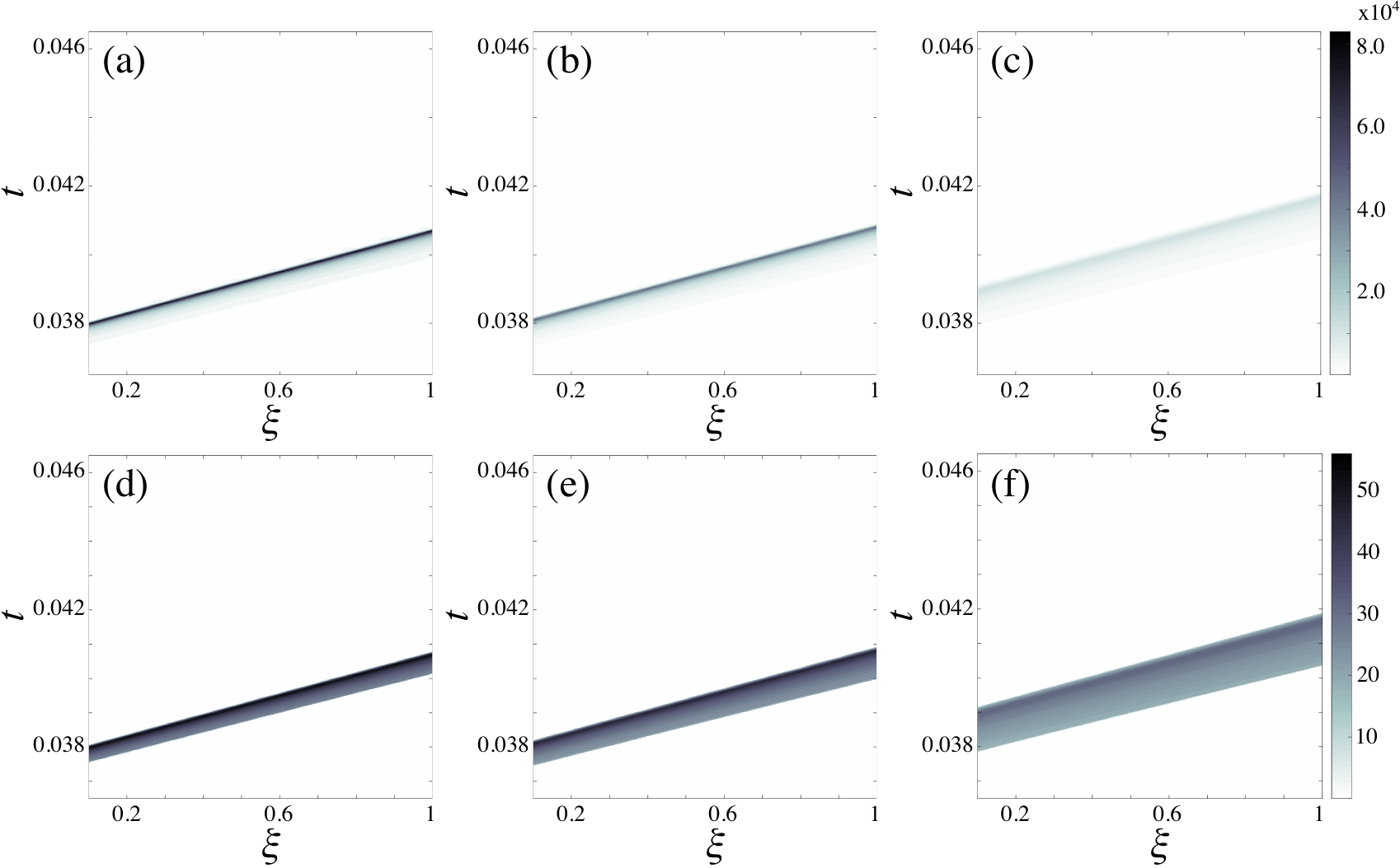} 
     \caption{Dependence of $\lambda_{max}$ (a-c) and $\mathrm{k}_{max}$ (d-f) on $\xi$ and  $\beta$, for $\beta=0.005$ (a,d), $\beta=0.05$ (b,e), and $\beta=0.5$ (c,f).}
   \label{fig:disprel2d}
\end{figure*}

We have additionally computed the values of $\lambda_{max}$ and $\mathrm{k}_{max}$ for different values of $\xi$ and $\beta$. The results are summarized in Fig.~\ref{fig:disprel2d}.

The first thing to be noticed is that $\lambda_{max}$ is significantly different from zero only in a narrow band, the smaller the value of $\beta$ the narrower the band, see Figs.~\ref{fig:disprel2d}(a-c). This can also be seen in Fig.~\ref{fig:evollambdamax}, where $\lambda_{max}$ is different from zero only in a narrow peak. Additionally, this band has a clear slope. This slope is of course related with coherency, higher values of $\xi$ imply a higher importance of the elastic energy, which is more important near the interface. These coherency strains delay phase separation, since the concentration needs to increase in order for the chemical energy to overcome the strain energy. Larger values of the flux parameter $F$ would bring phase separation to earlier times and also change this slope, since the necessary buildup of concentration would take less time. Note also that the peak value of $\lambda _{max}$ increases with $\xi$, albeit slightly. Similarly, the width of the time interval where $\lambda_{max}$ is significantly larger than zero increases with $\xi$, which can be more clearly appreciated in the plots of $k_{max}$, Figs.~\ref{fig:disprel2d}(d-f).

The effect of $\beta$ is also clearly shown on Fig.~\ref{fig:disprel2d}. Increasing $\beta$ decreases the peak value of $\lambda_{max}$ for all values of $\xi$, and at the same time widens the peak of the instability. Nevertheless, one effect does not compensate for the other, since the integral of $\lambda _{max}$ in the instability region is much smaller for the $\beta=0.5$ case than for the other two. The integral corresponds to an upper bound for the logarithm of the amplification of any perturbation, and hence we can conclude that the $\beta=0.5$ case is more stable in any case in the linear regime. 

The increase of  $\beta$ also delays the instability, as it had been anticipated before. The positions of the peak in the $\xi=0.1$ case are $t_{peak}=0.0380$, $t_{peak}=0.0381$, and $t_{peak}=0.0390$, for the cases with $\beta = 0.005$, $\beta=0.05$, and $\beta=0.5$, respectively.

The most unstable mode $\mathrm{k}_{max}$ follows a similar dependence with time as $\lambda_{max}$, as expected from Fig.~\ref{fig:evollambdamax}. It shows a weak dependence on $\xi$ along the peak, similarly to $\lambda_{max}$, and it raises much faster from the onset than $\lambda_{max}$, which explains the thicker band represented in  Fig.~\ref{fig:disprel2d}. 

\subsection{Instability of the receding front}

In this section we consider a fully phase separated layer, on which a negative flux ($F<0$) drives the interface between the lithiated and nonlithiated phases towards the absorption boundary. This receding interface in the case without elasticity is known to be unstable, in accordance with the well-known correspondence with the Hele-Shaw problem in the sharp-interface limit \cite{Pego1989}.

In our case, the stability  in the system corresponding to the sharp-interface limit has also been studied for the $\beta=0$ case \cite{Leo1989,Onuki1991}. In a previous article  \cite{Meca2016b} the authors have derived the sharp interface limit for the complete model, the main results are described in Appendix \ref{appendix}. We obtain the following dispersion relation for perturbations of the sharp interface:

\begin{equation}
\lambda=-\mathrm{k}\frac{F+2 I \mathrm{k}^2+Z \mathrm{k}}{1+2I\beta \mathrm{k}},\label{eq:growthratea}
\end{equation}
see Appendix \ref{appendix} for the definitions of $Z$ and $I$ and the details of the derivation, which is novel for the $\beta>0$ case. Inspection of Eq.~\eqref{eq:growthratea} reveals that the $F<0$ case will in general be unstable.

We can compare the dispersion relation obtained obtained with exactly the same procedure outlined in the previous sections with Eq.~\eqref{eq:growthratea}. This comparison, which should be accurate for a large enough system, fulfils a double purpose. On the one hand, it allows us to validate our results, since the two dispersion relations are derived in two exceedingly different ways. On the other hand, it allows us to test the convergence of the system with the value of $\varepsilon$.

In order to generate a receding interface we let evolve the system starting with completely depleted layer, and reverse the sign of $F$ at $t=0.2$, when the front is approximately in the middle of the layer. Then the dispersion relation and the eigenvalues are computed at $t=0.225$, at which point the transient corresponding to the sign reversal has decayed sufficiently. The layer is thicker than in the previous case, with a thickness of $2$, to facilitate the comparison with the unbounded case. The reversal of $F$ can be accomplished for the system at hand by stopping the driving current and connecting the electrode to a load.

\begin{figure}[htbp] 
   \centering
   \includegraphics[width=\columnwidth]{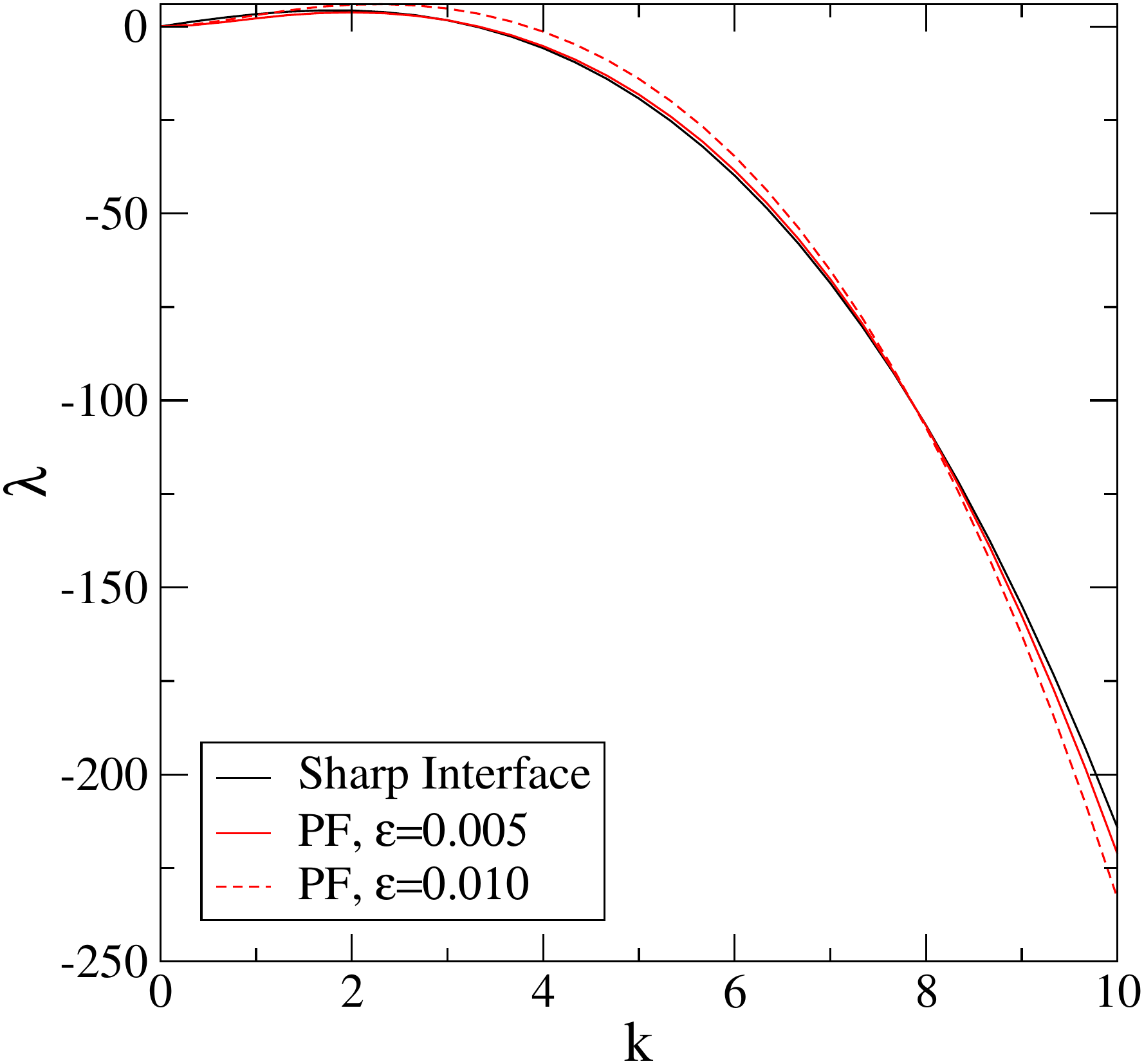} 
   \caption{Comparison of the phase field (PF) instability (largest eigenvalue) for the receding interface at $t=0.225$, with the sharp interface growth rate, Eq.~\eqref{eq:growthrate}. Parameters are $F=-4$, $\beta=0.05$ and $\xi=0.1$.}
   \label{fig:comdisp}
\end{figure}

The comparison (Fig.~\ref{fig:comdisp}) shows that the two methods give indeed very similar results, with a clear improvement as $\varepsilon$ is decreased. This good agreement is surprising, given that Eq.~\eqref{eq:growthratea} is derived for an unbounded system in the steady state, whereas the phase-field simulations are for a bounded system (albeit with a size that is the double of the previous section) that is in a transient state. This makes this good agreement even more remarkable. Nevertheless, the results show that the results are not so good for smaller $\mathrm{k}$, what we assume is an effect of the boundary conditions, and similarly $\varepsilon$ dependence is larger for large $\mathrm{k}$, which again is to be expected since these modes correspond to smaller wavelengths.

\begin{figure}[htbp] 
   \centering

   \includegraphics[width=\columnwidth]{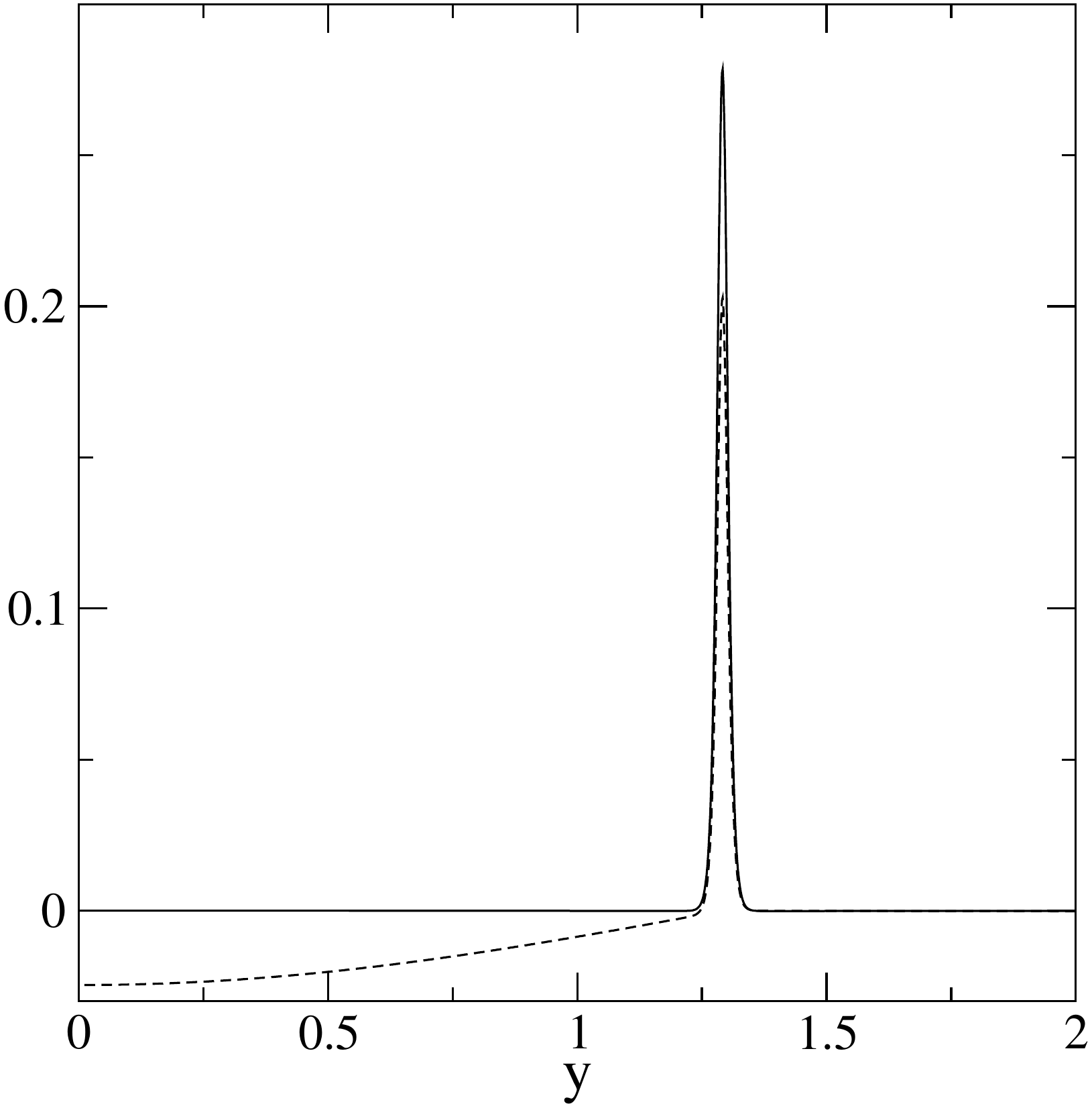} 
   \caption{Instability for $\beta=0.05$ and $\xi=0.1$. Most unstable eigenvectors at $t=0.225$ and $\mathrm{k}=1.94$. The solid and the dashed lines correspond respectively to the most unstable and the second most unstable eigenvectors.}
   \label{fig:evecreceding}
\end{figure}

The eigenvectors corresponding to the most unstable eigenvalues at $t=0.225$ have also been computed for $k_{max}=1.94$, (Fig.~\ref{fig:evecreceding}). Results show that the eigenvector from the most unstable eigenvalue is zero almost everywhere, except in the vicinity of the interface. On the one hand, this is to be expected, since the instability, which is akin to the Mullins-Sekerka instabiilty, is localized a the interface. On the other hand, this result is surprising, since we are treating the instabilities as a bulk phenomenon and we have obtained this localization in a natural way. In Fig.~\ref{fig:evecreceding} the eigenvector corresponding to the second largest eigenvalue, which is negative, is also on display. This eigenvector is not completely localized, but rather extends into the depleted part of the layer. This scenario is again very similar to the one shown in Fig~\ref{fig:disprel}, where only the eigenvector of the positive eigenvalue is localized.

Finally, note that this long wave instability would develop very slowly when compared with the instability related with phase separation described in the previous section. The inverse of $\lambda_{max}=3.12$ can be used as a proxy for the time for the development of the instability, which gives a time $t=0.32$, which is larger than all the times that have been considered in this work.

\section{Conclusion\label{Conclusion}}

In the present article we have used an unified approach to the study of the different instabilities that are present in the system of study. Through our study we have described a transient localized instability related with spinodal decomposition and found an unexpected connection with a Mullins-Sekerka-like instability that occurs in the phase separated case when the interface recedes. The present unified approach allows thus for the systematic and simultaneous study of instabilities that are typically not connected, allowing the mutual validation of the different techniques used to study them.

This article also incorporates the study of the role of kinetics on the transient instability, as well as on the receding front instability. While there are previous works that have derived equations similar to Eq.~\eqref{eq:growthratea}, such as \cite{Leo1989} and \cite{Onuki1991} this is to our knowledge the only derivation that incorporates the role of kinetics, thus we give a detailed account of the derivation in the appendix.

We have found the conditions under which the patterns formed in Figs.~\ref{fig:compbeta} and \ref{fig:compxi} develop, and have characterized the instability as a transient one. Nevertheless, our approach based in the linear regime has limitations, as exemplified by the case $\beta=0.5$, $\xi=1.0$, that according to our analysis should be less unstable, but give in fact a pattern that lasts longer in time, as shown in Fig. \ref{fig:compxi}.

Since this localized instability is transient, the
linearised problem has coefficients that are time dependent and non-uniform
in space and hence the variables cannot be separated. A common approach
\cite{Meca2010,lick_instability_1965,currie_effect_1967, edmonstone_surfactant-induced_2005,warner_unstable_2002,foster_stability_1965, munch_impact_2011, doumenc_coupling_2005,touazi_simulation_2010} used also in this paper is to ``freeze'' time (only) in the coefficients and then proceed with a traditional
separation of variables ansatz. This yields exponential evolution in time at
a rate that is determined by the solution of a spatial eigenvalue problem.
The question is to determine when this method is accurate. Moreover, the
obtained rate depends on the time at which the coefficients are frozen and
hence may lead to different results at different times. In particular, a
system may change from stable to unstable or vice-versa as the coefficients
are taken for progressively later times, and as is the case here, may be
unstable only for a limited period of time.

To incorporate the effect of the slowly changing coefficients,
a multiple scales ansatz can be used, see for example
\cite{hennessy-multiple-scale-2013, dziwnik-stability-2014} and references
in particular in \cite{hennessy-multiple-scale-2013}. This analysis reveals
two key conclusions: First, that the log of
the amplification of each mode is given by the integral of the eigenvalue
in time; and secondly, that this approximation is the leading order
contribution if the eigenvalue multiplied with the time scale over which
the coefficients change is large. In Fig.~\ref{fig:evollambdamax}, the peak
of the eigenvalue times the time over which it changes is indeed large,
so the the condition is satisfied. Then, the amplification can be estimated
by integrating the eigenvalue obtained from the frozen mode analysis, and
then exponentiating the result. Since the top eigenvalue changes sign,
we obtain a largest amplification after which the instability subsided.
In \cite{dziwnik-stability-2014} it was shown how the dominant mode can
be obtained by finding, at each time, the wave number with the largest
amplification. This is not the value $k_{\text{max}}$ that is obtained in this
paper, but the latter may be enough to indicate basic trends. A more detailed
investigation that determines the different time scales analytically and
their impact on the amplification of perturbations will be left to future work.

Finally, the scenario studied here in detail is relevant for applications where the flux $F$ is high enough, in the limit of small $F$ we obtain the scenario described in  Ref.~\cite{Tang2012}. One can thus reach that scenario from the one described here through the continuous dependence on $F$. We note, that the fact that the system is driven changes its behaviour dramatically, from the nature of the localization of the concentration to the finite $\mathrm{k}$ of the first instability, as opposed to a purely long-wavelength, spinodal-decomposition-like instability. The characterization of this transition from a concentration-dominated to an elastic-dominated instability is currently receiving our attention and can also be studied with the same model, but it is out of the scope of the present work.

\bibliography{LiSiModeling}
\bibliographystyle{abbrv}


\appendix

\section{Instability of the sharp-interface model\label{appendix}}

In this appendix we detail the instability of the sharp interface limit of Eqs.~\eqref{eq:ndimeqs} as computed by Meca et al. \cite{Meca2016b}. The equations for the chemical potential and the stress read as follows:

\begin{subequations}\label{eq:sharpinterface}
\begin{align}
\lapla{\mu}_0&=0,\label{eq:chempoteq}\\
\nabla\cdot{\boldsymbol{\sigma}}_0&=0,
\end{align}
together with the constitutive relation for stress:
\begin{align}
\sigma_{ij,0}=&2{G^{\pm}} \left(\epsilon_{ij,0}-\epsilon^{0,\pm}_{ij}\right)\nonumber\\
&+\frac{2\nu}{1-2\nu}{G^{\pm}}\left(\epsilon_{kk,0}-\epsilon^{0,\pm}_{kk}\right)\delta_{ij}, \label{eq:conststress}
\end{align}
where $G^{\pm}=G(c_0^\pm)$ and $\epsilon^{0,\pm}_{ij}=h(c_0^\pm)$ are constants. The $\pm$ superindex represents the values at the interface for both regions, the lithiated ($\Omega ^+$) and the amorphous silicon phase ($\Omega ^-$). These values have to be understood as liimits. The specific values of $G^{\pm}$ and $h(c_0^\pm)$ are
\begin{align}
G^\pm&=\left\{
\begin{array}{lr}
1& \mathbf{r}\in \Omega^-\\
\displaystyle{\frac{E_{\ce{Li_xSi}}}{E_{\ce{Si}}}}& \mathbf{r}\in \Omega^+
\end{array}
\right. , \nonumber\\
\eps^{0,\pm}_{ij}&=\left\{
\begin{array}{lr}
\displaystyle{h(c^-)\delta_{ij}=0}& \mathbf{r}\in \Omega^-\\
\displaystyle{h(c^+)\delta_{ij}=\delta _{ij}}& \mathbf{r}\in \Omega^+
\end{array}
\right. .
\end{align}

Relation \eqref{eq:conststress} can be inverted to yied
\begin{equation}
\epsilon_{ij,0}=\epsilon^{0,\pm}_{ij}+\frac{1}{2G^\pm}\sigma_{ij,0}-\frac{1}{2G^\pm}\frac{\nu}{1+\nu}\delta_{ij}\sigma_{kk}\label{eq:findstrain}
\end{equation}
for the strain tensor. This relation is explicitly used below.

Similarly, from the plane strain approximation the value of $\sigma _{zz,0}$ can be computed as follows:
\begin{equation}
\sigma _{zz,0}=-2(1+\nu)G^{\pm}\epsilon_{zz}^{0, \pm}+\nu\left(\sigma_{xx,0}+\sigma_{yy,0}\right). \label{eq:findzz}
\end{equation}

The boundary conditions {at the free boundary} for the elasticity equation correspond to continuity for the elastic field and for the tractions {across the interface}:
\begin{align}
{\mathbf{u}}_0^+&={\mathbf{u}}_0^-, \label{eq:contu}\\
\mathbf{n}\cdot{\boldsymbol{\sigma}}_0^+&=\mathbf{n}\cdot{\boldsymbol{\sigma}}_0^-. \label{eq:contsigma}
\end{align}
For the chemical potential equation we have at the interface away from the {absorption} boundary:
\begin{align}
{\mu}_0^\pm({c}_0^+-{c}_0^-)=&-\left(\beta\,v_n+\mathcal K\right)I \nonumber\\
&+\frac{\xi}{2}\left[{\sigma}_{ij,0}^+\left({\epsilon}_{ij,0}^+-\delta_{ij}h({c}_0^+)\right)\right. \nonumber\\
&\left.-{\sigma}_{ij,0}^-\left({\epsilon}_{ij,0}^--\delta_{ij}h({c}_0^-)\right)\right]\nonumber\\
&-\xi\,{\sigma}_{ij,0}^+\left({\epsilon}_{ij,0}^+-{\epsilon}_{ij,0}^-\right) \label{eq:mueqsummary},
\\ \nn\\
\left({c}_0^+-{c}_0^-\right)v_n=&-\left(\partial_r{\mu}_0^+-\partial_r{\mu}_0^-\right)\label{eq:conssummary},
\end{align}
where $I=\int_{0}^{1}\sqrt{2 f(\phi)}\,d\phi$. The conditions at the substrate are
\begin{align}
\partial_y{\mu}_0|_{y=0}&=0,\label{eq:BVsummarys}\\
\mathbf{u}|_{y=0}&=0, \label{eq:BVsummary2s}
\end{align}
and at the absorption boundary we have:
\begin{align}
\partial_y{\mu}_0|_{y=1}&=F,\label{eq:BVsummary}\\
\sigma_{iy,0}|_{y=1}&=0, \quad i=x,y.\label{eq:BVsummary2}
\end{align}
\end{subequations}
At the triple junctions the angle is $\alpha=\pi/2$.

This systems admits a one-dimensional travelling-wave solution, with the interface located at $y_I=-F t$. All of the components of the strain tensor are zero except for $\epsilon_{yy,0}$, which reads
\begin{equation}
\epsilon_{yy,0}=\left\{
\begin{array}{lr}
0& y<y_I\\
\displaystyle{\frac{1+\nu}{1-\nu}}& y>y_I
\end{array}
\right. ,
\end{equation}
which implies that $u_{x,0}=u_{z,0}=0$ and therefore
\begin{equation}
u_{y,0}=\left\{
\begin{array}{lr}
0& y<y_I\\
\displaystyle{\frac{1+\nu}{1-\nu}(y+Ft)}& y>y_I
\end{array}
\right. .
\end{equation}

Similarly, the value of all components of stress is zero except for $\sigma_{xx,0}$ and $\sigma_{zz,0}$, they are both equal to
\begin{equation}
\sigma_{xx,0}=\left\{
\begin{array}{lr}
0& y<y_I\\
\displaystyle{-2\frac{E_{\ce{Li_xSi}}}{E_{\ce{Si}}}\frac{1+\nu}{1-\nu}}& y>y_I
\end{array}
\right. .
\end{equation}

Finally, we have for the chemical potential
\begin{equation}
\mu_{y,0}=\left\{
\begin{array}{lr}
\displaystyle{IF\beta + R\xi \xi}& y<y_I\\
\displaystyle{F(y+Ft) + R}& y>y_I
\end{array}
\right. ,
\end{equation}
with 
\begin{equation}
R=\frac{E_{\ce{Li_xSi}}}{E_{\ce{Si}}}\frac{1+\nu}{1-\nu}, 
\end{equation}
which is obviously continuous. Notice that in all the previous cases a temporal translation is enough to give the appropriate initial conditions, and that this travelling wave fulfils all of the boundary conditions at the interface and on the outer boundaries.

\subsection{Stability of the one-dimensional solution}

The previously described solution can be perturbed in order to asses its stability.  We will use an Airy stress function in order to treat in a unified way the displacement vector and the strain and stress tensors.
\begin{equation}
\sigma _{xx}=\partial ^2 _y\phi, \;\;\;\; \sigma _{yy}=\partial ^2 _x\phi, \;\;\;\; \sigma _{xy}=-\partial ^2 _{xy}\phi.
\end{equation}
It  can be proved that $\phi$ satisfies the biharmonic equation
\begin{equation}
\lapla\lapla \phi =0,\label{eq:biharmonic}
\end{equation}
as long as the elastic constants do not vary and there is a constant or linearly varying eigenstrain. Fields $\phi$ and $\mu$ are perturbed as follows:
\begin{subequations}
\begin{align}
\phi&=\phi_0+\varepsilon\phi_1,\\
\mu&=\mu_0+\varepsilon \mu_1,
\end{align}
\label{eq:expansion}
\end{subequations}
where $\varepsilon$ is a formal expansion parameter. We take $\phi_1$ and $\mu_1$ as periodic in the x direction, and assume an exponential dependence on time:
\begin{subequations}
\begin{align}
\phi_1&=e^{\lambda t}e^{i\mathrm{k}x}\Phi(y),\\
\mu_1&=e^{\lambda t}e^{i\mathrm{k}x}M(y).
\end{align}
\label{eq:perturbation}
\end{subequations}
Substituting \eqref{eq:expansion} and \eqref{eq:perturbation} into Eqs.~\eqref{eq:chempoteq} and \eqref{eq:biharmonic} linear ODEs are obtained that give the following general solution:
\begin{subequations}
\begin{align}
\Phi(y)&=(A_1^\pm+A_3^\pm y)e^{-\mathrm{k}y}+(A_2^\pm+A_4^\pm y)e^{\mathrm{k}y},\\
M(y)&=B_1 ^{\pm}  e^{-\mathrm{k}y}+B_2 ^{\pm}  e^{\mathrm{k}y},
\end{align}
\end{subequations}
where $A^\pm _i$ and $B^\pm _i$ are constants, and the $\pm$ superindices denote both sides of the interface. The position of the interface is similarly perturbed:
\begin{equation}
\Upsilon(x) = y_I(t)+\varepsilon \Upsilon _1 e^{\lambda t} e^{i\mathrm{k}x},
\label{eq:perturbationint}
\end{equation}
where $\Upsilon _1$ is a constant. From the previous equation we obtain the form of the normal vector:
\begin{align}
\mathbf{n} &=\frac{1}{\sqrt{(\partial_x \Upsilon)^2+1}}\left(
\begin{array}{c}
-\partial_x \Upsilon\\
1
\end{array}
\right)\nonumber\\
&=
\left(
\begin{array}{c}
0\\
1
\end{array}
\right)
+
\varepsilon
\left(
\begin{array}{c}
i\mathrm{k} \Upsilon _1 e^{\lambda t} e^{i\mathrm{k}x}\\
0
\end{array}
\right)
+O\left(\varepsilon ^2\right).
\end{align}

The perturbations \eqref{eq:perturbation} and \eqref{eq:perturbationint} contain a total of 13 constants. They can be found from the boundary conditions \eqref{eq:contu}, \eqref{eq:contsigma}, \eqref{eq:mueqsummary}, \eqref{eq:conssummary}, \eqref{eq:BVsummarys}, \eqref{eq:BVsummary2s}, \eqref{eq:BVsummary}, and \eqref{eq:BVsummary2}, which also sum 13 conditions.

The introduction of the perturbations in the equations will lead to a homogeneous system of 13 equations. They would give rise to a homogeneous system, and requiring that there exists a solution other than the trivial results in a dispersion relation that gives the growth rate $\sigma$ as a function of the wavenumber $\mathrm{k}$.

\subsubsection{Solution of the unbounded case}
In this case we can use a travelling wave ansatz for the perturbation, by changing $y\rightarrow \tilde y + y_I$, such that $y=y_I$ implies $\tilde y=0$ (we drop the tilde signs from now on). The equations are invariant under this transformation, and the equations are considerably simplified. The solutions are the same, but imposing that the perturbations are finite at infinity gives directly:
\begin{equation}
A^-_1=A^-_3=A^+_2=A^+_4=B^-_1=B^+_2=0,
\end{equation}
which simplifies the equations considerably. From the conservation condition \eqref{eq:conssummary} we obtain
\begin{align}
-F+\varepsilon \lambda \Upsilon_1 e^{\lambda t} e^{i\mathrm{k}x}&=\nonumber\\
&\hspace*{-0.95cm}-F-\varepsilon \mathrm{k}\left(-B_1^+ -B_2^-\right)e^{\lambda t} e^{i\mathrm{k}x},
\end{align}
and hence
\begin{equation}
\lambda \Upsilon_1 = \mathrm{k}\left(B_1^+ +B_2^-\right). \label{eq:1cons}
\end{equation}
In order to write the form of the local equilibrium condition  \eqref{eq:mueqsummary}, we need the explicit form of the stress and strain tensors. For $\mathbf{r}\in \Omega_+$ we have that
\begin{subequations}
\begin{align}
\sigma_{xx}&=-2\frac{E_{\ce{Li_xSi}}}{E_{\ce{Si}}}\frac{1+\nu}{1-\nu}\nonumber\\
&+\varepsilon \left[\mathrm{k}^2 A_1^+ + \left(\mathrm{k}^2y-2\mathrm{k}\right)A_3^+\right]e^{\lambda t}e^{i\mathrm{k}x}e^{-\mathrm{k}y}\\
\sigma_{yy}&=-\varepsilon \mathrm{k}^2\left(A_1^+ + A_3^+y\right)e^{\lambda t}e^{i\mathrm{k}x}e^{-\mathrm{k}y}\\
\sigma_{xy}&=-\varepsilon i \mathrm{k}\left(A_3^+-\mathrm{k}A_1^+-\mathrm{k}A_3^+y\right)e^{\lambda t}e^{i\mathrm{k}x}e^{-\mathrm{k}y}
\end{align}
\end{subequations}

The value of $\sigma _{zz}$ can be computed from the previous equations by using Eq.~\eqref{eq:findzz}, which results in
\begin{equation}
\sigma _{zz}=-2\frac{1+\nu}{1-\nu}\frac{E_{\ce{Li_xSi}}}{E_{\ce{Si}}}-2\nu \mathrm{k} \varepsilon A_3^+ e^{\lambda t}e^{i\mathrm{k}x}e^{-\mathrm{k}y}.
\end{equation}
Therefore,
\begin{equation}
\sigma _{kk}=-4\frac{1+\nu}{1-\nu}\frac{E_{\ce{Li_xSi}}}{E_{\ce{Si}}}-2(1+\nu) \mathrm{k} \varepsilon A_3^+ e^{\lambda t}e^{i\mathrm{k}x}e^{-\mathrm{k}y}.
\end{equation}

By using the previous result and Eq.~\eqref{eq:findstrain}, the non-zero components of the strain tensor can be computed
\begin{subequations}
\begin{align}
\epsilon_{xx}&=\frac{\varepsilon \left[\mathrm{k}^2 A_1^+ + \left(\mathrm{k}^2y-2(1-\nu)\mathrm{k}\right)A_3^+\right]}{2G^+}e^{\lambda t + i\mathrm{k}x-\mathrm{k}y}\\
\epsilon_{yy}&=\frac{1+\nu}{1-\nu}\nonumber\\
&-\frac{\varepsilon \left[\mathrm{k}^2A_1^+ + (\mathrm{k}^2y-2\nu k)A_3^+\right]}{2G^+}e^{\lambda t + i\mathrm{k}x -\mathrm{k}y},\\
\epsilon_{xy}&=-\frac{\varepsilon i \mathrm{k}\left(A_3^+-\mathrm{k}A_1^+-\mathrm{k}A_3^+y\right)}{2G^+}e^{\lambda t + i\mathrm{k}x-\mathrm{k}y}.
\end{align}
\end{subequations}

The displacement functions can be obtained by integration (by using the definition of the shear stress as a compatibility condition),
\begin{subequations}
\begin{align}
u_x&=\int \epsilon _{xx} dx + Ay+x_0,\\
u_y&=\int \epsilon _{yy} dy - Ax+y_0,
\end{align}
\end{subequations}
i.e. the displacements associated with strain plus an infinitesimal rotation of angle $A$ and a translation $(x_0,y_0)$, two strainless transformations. Since both of these additions imply a displacement at infinity we can safely ignore them. The final result is then
\begin{subequations}
\begin{align}
u_x&=\frac{-i \varepsilon \left[\mathrm{k} A_1^+ + \left(\mathrm{k}y-2(1-\nu)\right)A_3^+\right]}{2G^+}e^{\lambda t +i\mathrm{k}x-\mathrm{k}y},\\
u_y&=\frac{1+\nu}{1-\nu}y\nonumber\\
&+\frac{\varepsilon \left[\mathrm{k}A_1^+ + (\mathrm{k}y+1-2\nu )A_3^+\right]}{2G^+}e^{\lambda t+i\mathrm{k}x-\mathrm{k}y}.
\end{align}
\end{subequations}
Of course the displacements are real and we will only retain the real part in the end.

For $\mathbf{r}\in \Omega_-$ we have that
\begin{subequations}
\begin{align}
\sigma_{xx}&=\varepsilon \left[\mathrm{k}^2 A_2^- + \left(\mathrm{k}^2y+2\mathrm{k}\right)A_4^-\right]e^{\lambda t}e^{i\mathrm{k}x}e^{\mathrm{k}y}\\
\sigma_{yy}&=-\varepsilon \mathrm{k}^2\left(A_2^- + A_4^-y\right)e^{\lambda t}e^{i\mathrm{k}x}e^{\mathrm{k}y}\\
\sigma_{xy}&=-\varepsilon i \mathrm{k}\left(A_4^- +\mathrm{k}A_2^- +\mathrm{k}A_4^-y\right)e^{\lambda t}e^{i\mathrm{k}x}e^{\mathrm{k}y}
\end{align}
\end{subequations}

The value of $\sigma _{zz}$ and $\sigma_{kk}$ can likewise be found:
\begin{align}
\sigma _{zz}&=2\nu \mathrm{k} \varepsilon A_4^- e^{\lambda t}e^{i\mathrm{k}x}e^{-\mathrm{k}y},\\
\sigma _{kk}&=2(1+\nu) \mathrm{k} \varepsilon A_4^- e^{\lambda t}e^{i\mathrm{k}x}e^{-\mathrm{k}y}.
\end{align}

Also the non-zero strain elements:
\begin{subequations}
\begin{align}
\epsilon_{xx}&=\frac{\varepsilon \left[\mathrm{k}^2 A_2^- + \left(\mathrm{k}^2y+2(1-\nu)\mathrm{k}\right)A_4^-\right]}{2G^-}e^{\lambda t+i\mathrm{k}x+\mathrm{k}y},\\
\epsilon_{yy}&=-\frac{\varepsilon \left[\mathrm{k}^2A_2^- + (\mathrm{k}^2y+2k\nu)A_4^-\right]}{2G^-}e^{\lambda t+i\mathrm{k}x+\mathrm{k}y},\\
\epsilon_{xy}&=-\frac{\varepsilon i \mathrm{k}\left(A_4^- +\mathrm{k}A_2^- +\mathrm{k}A_4^-y\right)}{2G^-}e^{\lambda t+i\mathrm{k}x+\mathrm{k}y}.
\end{align}
\end{subequations}

Proceeding in the same way as before, we obtain the displacements
\begin{subequations}
\begin{align}
u_{x}&=\frac{-i \varepsilon \left[\mathrm{k} A_2^- + \left(\mathrm{k}y+2(1-\nu)\right)A_4^-\right]}{2G^-}e^{\lambda t+i\mathrm{k}x+\mathrm{k}y},\\
u_{y}&=-\frac{\varepsilon \left[\mathrm{k}A_2^- + (\mathrm{k}y-1+2\nu)A_4^-\right]}{2G^-}e^{\lambda t+i\mathrm{k}x+\mathrm{k}y}.
\end{align}
\end{subequations}

We can introduce the previous expressions for the displacement and the stress in Eqs.~\eqref{eq:contu} and \eqref{eq:contsigma}, and substitute $y=\varepsilon \Upsilon _1 e^{\lambda t} e^{i\mathrm{k}x}$. Retaining terms at $O(\varepsilon)$ we obtain
\begin{subequations}
\begin{align}
&\mathrm{k} A_1^+ -2(1-\nu)A_3^+-Q-1=0\\
&2R\Upsilon _1+\mathrm{k}A_1^+ + (1-2\nu )A_3^++Q_2=0\\
&-2R\Upsilon _1 - A_3^++\mathrm{k}A_1^++A_4^- +\mathrm{k}A_2^- =0\\
&A_1^+ -A_2^- =0
\end{align}
\label{eq:elastcont}
\end{subequations}
with 
\begin{equation}
Q_1=\frac{E_{\ce{Li_xSi}}}{E_{\ce{Si}}}\left[\mathrm{k} A_2^- +2(1-\nu)A_4^-\right]
\end{equation}
\begin{equation}
Q_2=\frac{E_{\ce{Li_xSi}}}{E_{\ce{Si}}}\left[\mathrm{k}A_2^- -(1-2\nu)A_4^-\right]
\end{equation} 
We obtain two additional conditions from Eq.~\eqref{eq:mueqsummary} 
\begin{subequations}
\begin{align}
F\Upsilon_1+B_1^+ &= -\left(\mathrm{k}^2+\lambda \beta \right)I \Upsilon_1 \\
&+\xi\frac{1+\nu}{1-\nu}\left\{\mathrm{k}^2 A_1^+ 
-Q_1\right\}, \nonumber\\
B_2^- &= -\left(\mathrm{k}^2+\lambda \beta \right)I \Upsilon_1 \\ 
&+\xi\frac{1+\nu}{1-\nu}\left\{\mathrm{k}^2 A_1^+ 
-Q_1\right\}. \nonumber
\end{align}
\label{eq:loceq}
\end{subequations}

Eqs.~\eqref{eq:1cons}, \eqref{eq:elastcont} and \eqref{eq:loceq} constitute then the expected homogeneous system of 7 equations with seven unknowns, $A_1^+$, $A_2^-$, $A_3^+$, $A_4^-$, $B_1^+$, $B_2^-$ and $\Upsilon_1$. Imposing that the determinant is zero to obtain other solutions than the trivial leads to the following expression for the growth rate $\lambda$:
\begin{equation}
\lambda=-\mathrm{k}\frac{F+2 I \mathrm{k}^2+Z \mathrm{k}}{1+2I\beta \mathrm{k}},\label{eq:growthrate}
\end{equation}
where $Z$ is a constant:
\begin{equation}
Z=8\xi\frac{\frac{E_{\ce{Li_xSi}}}{E_{\ce{Si}}}\left(1+\frac{E_{\ce{Li_xSi}}}{E_{\ce{Si}}}\right)(1+\nu)^2 }{\left(3-4\nu+\frac{E_{\ce{Li_xSi}}}{E_{\ce{Si}}}\right)(1-\nu)},
\end{equation}
which contains all the elastic constants. Clearly, we recover the expected Mullins-Sekerka dispersion relation (augmented with the kinetic term) in the limit $\xi\rightarrow0$, and the constant $Z>0$, and hence it will have an stabilizing effect.



\end{document}